\documentclass[a4paper,12pt]{article}
\usepackage{eurosym}
\usepackage[dvips]{graphicx}
\usepackage{amsfonts}
\usepackage{mathrsfs}
\usepackage{graphicx} 
\usepackage{epsfig}
\usepackage{amsmath, amsthm}
\usepackage{amssymb}

\newcommand{\ba}{\begin{array}}
\newcommand{\ea}{\end{array}}

\newcommand{\beq}{\begin{equation}}
\newcommand{\eeq}{\end{equation}}
\newcommand{\ben}{\begin{enumerate}}
\newcommand{\een}{\end{enumerate}}
\newcommand{\bit}{\begin{itemize}}
\newcommand{\eit}{\end{itemize}}

\begin{document}
\title{Generalized Heisenberg Ferromagnet  type  Equation and Modified Camassa-Holm Equation: Geometric  Formulation,  Soliton Solutions  and  Equivalence}
\author{Zhanar Umurzakhova\footnote{Email: zumurzakhova@gmail.com}, \, Tolkynay Myrzakul\footnote{Email: kryesmakhanova@gmail.com}, \, Kuralay Yesmakhanova\footnote{Email: kryesmakhanova@gmail.com}, \\Gulgassyl Nugmanova\footnote{Email: gnnugmanova@gmail.com}, \, Shynaray Myrzakul\footnote{Email: srmyrzakul@gmail.com}\,  and Ratbay Myrzakulov\footnote{Email: rmyrzakulov@gmail.com}\\
\textsl{Eurasian International Center for Theoretical Physics and} \\ { Department of General \& Theoretical Physics}, \\ Eurasian National University,
Nur-Sultan, 010008, Kazakhstan
}
\date{}
\maketitle

\begin{abstract}
We study the integrability and equivalence   of a generalized Heisenberg ferromagnet-type equation (GHFE). The different forms of this equation as well as its reduction are  presented. The Lax representation (LR) of the equation is obtained. We observe that the geometrical and gauge equivalent counterpart of the GHFE is the modified Camassa-Holm equation (mCHE) with an arbitrary parameter $\kappa$. Finally, the 1-soliton solution of the GHFE is obtained.
\end{abstract}


\section{Introduction}
This work continues  our research of Lax-integrable (i.e., admitting Lax pairs with non-vanishing
spectral parameter) generalized Heisenberg ferromagnet  type equations in 1+1 dimensions related with Camassa-Holm type equations (see, e.g., \cite{1907.10910}-\cite{1910.13281} and the references therein). In the theory of integrable systems (soliton theory) an important role plays the so-called gauge and geometrical  equivalence between two integrable equations. The well-known  classical example such equivalences is the  (gauge and geometrical) equivalence between the Heisenberg ferromagnet equation (HFE)
\begin{eqnarray}
iA_{t}+\frac{1}{2}[A,A_{xx}]=0 \label{1}
\end{eqnarray}
and the nonlinear Schr\"odinger equation (NLSE)
\begin{eqnarray}
iq_{t}+q_{xx}+2|q|^{2}q=0, \label{2}
\end{eqnarray}
where $q$ is a complex function and 
\begin{eqnarray}
A=\left(\ba{cc}A_{3}&A^{-}\\A^{+}&-A_{3}\ea\right), \quad A^{2}=I, \quad {\bf A}=(A_{1},A_{2},A_{3}), \quad {\bf A}^{2}=1. \label{3}
\end{eqnarray}

In this paper, we study the Generalized Heisenberg ferromagnet - type equation (GHFE), namely, the so-called  M-CXII equation (M-CXIIE) and its relation with  the modified Camassa-Holm  equation (mCHE)
\begin{eqnarray}
m_{t}+(m(u^2-u^2_{x}))_{x}+\kappa u_{x}&=&0,\\
 m-u+u_{xx}&=&0,
\end{eqnarray}
where $u=u(x,t)$ is a real-valued function, $\kappa=const.$  The mCHE (4)-(5)  was proposed  by Fuchssteiner \cite{fpd1996} and Olver and Rosenau \cite{orpre1996}. It was obtained from the two-dimensional Euler equations, where the variables $u(x,t)$ and $m(x,t)$ represent,
respectively, the velocity of the fluid and its potential density \cite{qjmp2006}. Some properties of the mCHE  and other related equations were studied in  \cite{gloqcmp2013} - \cite{0805.4310}. 

This paper is organized as follows. In Section 2, we give the M-CXIIE,   its different forms, its Lax representation (LR) and a reduction.  Geometric formulation of this equation in terms of space and plane curves is presented in Section 3. Using this geometric formalism, the geometrical equivalence between the M-CXIIE and the mCHE is established.   Section 4 is devoted to the mCHE. Gauge equivalence between the M-CXIIE and the mCHE and the relation between their solutions we have established  in Section 5.   The 1-soliton solution of the M-CXIIE is obtained in Section 6. We discuss and conclude our results in Section 7.

\section{The generalized Heisenberg ferromagnet  type equation}
\subsection{Equation}
There are exist several integrable and nonintegrable  GHFE  (see, e.g., \cite{R13}-\cite{zh1}). In this paper, we consider one of such GHFE, namely, the so-called M-CXIIE. This equation has the form
\begin{eqnarray}
[A,A_{xt}]+(\phi[A,A_{x}])_{x}+\frac{\kappa u_{x}}{m}[A,A_{x}]+\frac{4\alpha_{0}}{\beta^{2}}A_{x}=0, \label{6}
\end{eqnarray}
where $\phi=u^{2}-u^{2}_{x}$ and 
\begin{eqnarray}
A=\left(\ba{cc}A_{3}&A^{-}\\A^{+}&-A_{3}\ea\right), \quad A^{2}=I, \quad {\bf A}=(A_{1},A_{2},A_{3}), \quad {\bf A}^{2}=1. \label{7}
\end{eqnarray}
We can also write the  M-CXIIE in the following form
\begin{eqnarray}
A_{xt}+\phi A_{xx}+v_{1}A+v_{2}A_{x}+\frac{\alpha_{0}}{\beta^{2}}[A,A_{x}]=0, \label{8}
\end{eqnarray}
where
\begin{eqnarray}
v_{1}=0.5\{A_{t},A_{x}\}+\phi A_{x}^{2}=-2umI, \quad v_{2}=\phi_{x}+\frac{\kappa u_{x}}{m}. \label{9}
\end{eqnarray}
Finally let us present the vector form of the M-CXIIE. It  reads as
\begin{eqnarray}
{\bf A}_{xt}+\phi {\bf A}_{xx}+v_{1}{\bf A}+v_{2}{\bf A}_{x}+\frac{\alpha_{0}}{\beta^{2}}{\bf A}\wedge {\bf A}_{x}=0. \label{10}
\end{eqnarray}

\subsection{Lax representation}
The M-CXIIE (6) is integrable. Its LR is given by
\begin{eqnarray}
\Phi_x&=&U_{1}\Phi,\\
\Phi_t&=&V_{1}\Phi,
\end{eqnarray}
where
\begin{eqnarray}
U_{1}&=&\frac{\alpha_{0}-\alpha}{2}A+\frac{\lambda-\beta}{4\beta}[A,A_{x}], \\
V_{1}&=&(\omega-\omega_{0})A+\{\frac{(\lambda-\beta)u}{2m\beta^{2}\lambda}-\frac{(\lambda-\beta)\phi}{4\beta}\}[A,A_{x}]+\frac{u_{x}(\alpha\beta-\alpha_{0}\lambda)}{\beta^{2}\lambda m}
\end{eqnarray}
or
\begin{eqnarray}
U_{1}&=&\frac{\alpha_{0}-\alpha}{2}A+\frac{\lambda-\beta}{4\beta}[A,A_{x}], \\
V_{1}&=&(\omega-\omega_{0})A+\frac{(\lambda-\beta)}{4\beta}\{\frac{2u}{m\beta\lambda}-\phi\}[A,A_{x}]+\frac{u_{x}}{\beta m}\{\frac{\alpha}{\lambda}-\frac{\alpha_{0}}{\beta}\}A_{x}.
\end{eqnarray}
Here $\beta=const,  \quad \alpha_{0}=\alpha|_{\lambda=\beta}, \quad  \alpha=\sqrt{1-\frac{1}{2}\kappa \lambda^2}, \quad \omega_{0}=\omega|_{\lambda=\beta}$, 
\begin{eqnarray}
  \omega=\alpha\lambda^{-2}+0.5\alpha\phi, \quad u=\pm\beta^{-1}(1-\partial_{x}^{2})^{-1}\sqrt{0.5tr(A_{x})^{2}}
\end{eqnarray}
and
\begin{eqnarray}
  \phi=u^{2}-u^{2}_{x}, \quad m=\pm\beta^{-1}\sqrt{0.5tr(A_{x})^{2}}=u-u_{xx}=(1-\partial_{x}^{2})u.
\end{eqnarray}

\subsection{Reductions}
The M-CXIIE admits some reductions. For example in the case $\kappa=0$, the M-CXIIE reduces to the so-called  M-CXIE having the  form \cite{16}
\begin{eqnarray}
[A,A_{xt}]+(\phi[A,A_{x}])_{x}+\frac{4}{\beta^{2}}A_{x}=0.  \label{2}
\end{eqnarray}
This integrable GHFE was studied in \cite{16}. Its LR is given by
\begin{eqnarray}
\Phi_x&=&U_{2}\Phi,\\
\Phi_t&=&V_{2}\Phi,
\end{eqnarray}
where $(z=\phi+\lambda^{2}, z_{0}=\phi+\beta^{-2})$ and 
\begin{eqnarray}
U_{2}&=&\frac{\lambda-\beta}{4\beta}[A,A_{x}], \\
V_{2}&=&(z-z_{0})A+\{\frac{(\lambda-\beta)u}{2m\beta^{2}\lambda}-\frac{(\lambda-\beta)\phi}{4\beta}\}[A,A_{x}]+\frac{u_{x}(\beta-\lambda)}{\beta^{2} \lambda m}A_{x}.
\end{eqnarray}

\section{Integrable motion of  curves induced by the M-CXIIE}
The aim of this section is to present the geometric formulation of the M-CXIIE in terms of  curves and to find its geometrical equivalent counterpart.
\subsection{Integrable motion of space curves}
We start from the differential geometry of space curves. In this subsection, we  consider  the integrable motion of space curves induced by the  M-CXIIE.   As usual, let us consider a smooth space curve ${\bf \gamma} (x,t): [0,X] \times [0, T] \rightarrow R^{3}$ in $R^{3}$. Let  $x$ is the arc length of the curve at each time $t$.   In differential language, such curve is given by the    Frenet-Serret equation (FSE). The FSE and its temporal counterpart look like 
\begin{eqnarray}
\left ( \begin{array}{ccc}
{\bf  e}_{1} \\
{\bf  e}_{2} \\
{\bf  e}_{3}
\end{array} \right)_{x} = C
\left ( \begin{array}{ccc}
{\bf  e}_{1} \\
{\bf  e}_{2} \\
{\bf  e}_{3}
\end{array} \right),\quad
\left ( \begin{array}{ccc}
{\bf  e}_{1} \\
{\bf  e}_{2} \\
{\bf  e}_{3}
\end{array} \right)_{t} = G
\left ( \begin{array}{ccc}
{\bf  e}_{1} \\
{\bf  e}_{2} \\
{\bf  e}_{3}
\end{array} \right), \label{24} 
\end{eqnarray}
where ${\bf e}_{j}$ are the   unit tangent vector $(j=1)$,  principal normal vector $(j=2)$ and binormal vector $(j=3)$ which given by ${\bf e}_{1}={\bf \gamma}_{x}, \quad {\bf e}_{2}=\frac{{\bf \gamma}_{xx}}{|{\bf \gamma}_{xx}|}, \quad {\bf e}_{3}={\bf e}_{1}\wedge {\bf e}_{2}, $ 
respectively.
Here
\begin{eqnarray}
C&=&
\left ( \begin{array}{ccc}
0   & \kappa_{1}     & \kappa_{2}  \\
-\kappa_{1}  & 0     & \tau  \\
-\kappa_{2}    & -\tau & 0
\end{array} \right)=-\tau L_{1}+\kappa_{2}L_{2}-\kappa_{1}L_{3} \in so(3),\\
G&=&
\left ( \begin{array}{ccc}
0       & \omega_{3}  & \omega_{2} \\
-\omega_{3} & 0      & \omega_{1} \\
-\omega_{2}  & -\omega_{1} & 0
\end{array} \right)=-\omega_{1}L_{1}+\omega_{2}L_{2}-\omega_{3}L_{3}\in so(3),\label{3.26} 
\end{eqnarray}
where  $\tau$,  $\kappa_{1}, \kappa_{2}$ are the  "torsion",  "geodesic curvature" and  "normal curvature" of the curve, respectively; $\omega_{j}$ are some  functions. Note that $L_{j}$ are basis elements of $so(3)$ algebra and have the forms  
\begin{eqnarray}
L_{1}=
\left ( \begin{array}{ccc}
0   & 0     & 0  \\
0  & 0     & -1  \\
0    & 1 & 0
\end{array} \right), \quad L_{2}=
\left ( \begin{array}{ccc}
0       & 0  & 1 \\
0 & 0      & 0 \\
-1  & 0 & 0
\end{array} \right), \quad L_{3}=
\left ( \begin{array}{ccc}
0       & -1  & 0\\
1 & 0      & 0 \\
0  & 0 & 0
\end{array} \right).
\end{eqnarray}
 They satisfy  the following commutation relations
\begin{eqnarray}
[L_{1}, L_{2}]=L_{3}, \quad [L_{2}, L_{3}]=L_{1},  \quad [L_{3}, L_{1}]= L_{2}.
\end{eqnarray}
In the following, we need also in the basis elements of $su(2)$ algebra. They have the forms 
\begin{eqnarray}
e_{1}=
\frac{1}{2i}\left ( \begin{array}{cc}
0       & 1  \\
1 & 0
\end{array} \right), \quad e_{2}=
\frac{1}{2i}\left ( \begin{array}{cc}
0       & -i  \\
i & 0
\end{array} \right), \quad e_{3}=
\frac{1}{2i}\left ( \begin{array}{cc}
1      & 0  \\
0 & -1
\end{array} \right),
\end{eqnarray}
where the Pauli matrices have the form
\begin{eqnarray}
\sigma_{1}=
\left ( \begin{array}{cc}
0       & 1  \\
1 & 0
\end{array} \right), \quad \sigma_{2}=
\left ( \begin{array}{cc}
0       & -i  \\
i & 0
\end{array} \right), \quad \sigma_{3}=
\left ( \begin{array}{cc}
1      & 0  \\
 0& -1
\end{array} \right).
\end{eqnarray}
These elements satisfy the following commutation relations
\begin{eqnarray}
[e_{1}, e_{2}]=e_{3}, \quad [e_{2}, e_{3}]=e_{1},  \quad [e_{3}, e_{1}]= e_{2}.
\end{eqnarray}
Note that the Pauli matrices obey the following commutation relations
\begin{eqnarray}
[\sigma_{1}, \sigma_{2}]=2i\sigma_{3}, \quad [\sigma_{2}, \sigma_{3}]=2i\sigma_{1},  \quad [\sigma_{3}, \sigma_{1}]= 2i\sigma_{2}
\end{eqnarray}
or
\begin{eqnarray}
[\sigma_{i}, \sigma_{j}]=2i\epsilon_{ijk}\sigma_{k}.
\end{eqnarray}
The well-known  isomorphism between the Lie algebras $su(2)$ and $so(3)$ means the following  correspondence between their basis elements $L_{j}\leftrightarrow e_{j}$. Using this isomorphism let us construct the following two matrices
\begin{eqnarray}
U&=&-\tau e_{1}+\kappa_{2}e_{2}-\kappa_{1}e_{3}=-\frac{1}{2i}\left ( \begin{array}{cc}
\kappa_{1}      & \tau+i\kappa_{2}  \\
 \tau-i\kappa_{2} &-\kappa_{1}
\end{array} \right)=\left ( \begin{array}{cc}
u_{11}      & u_{12}  \\
 u_{21} &-u_{11}
\end{array} \right),\\
V&=&-\omega_{1}e_{1}+\omega_{2}e_{2}-\omega_{3}e_{3}=-\frac{1}{2i}\left ( \begin{array}{cc}
\omega_{3}      & \omega_{1}+i\omega_{2} \\
 \omega_{1}-i\omega_{2} &-\omega_{3}
\end{array} \right)=\left ( \begin{array}{cc}
v_{11}  & v_{12}  \\
 v_{21} &-v_{11}
\end{array} \right).\label{3.26} 
\end{eqnarray}
Hence we obtain
\begin{eqnarray}
\kappa_{1}&=&-2iu_{11}, \quad \kappa_{2}=-(u_{12}-u_{21}), \quad \tau=-i(u_{12}+u_{21}),\label{3.26} \\
\omega_{1}&=&-i(v_{12}+v_{21}), \quad \omega_{2}=-(v_{12}-v_{21}), \quad \omega_{3}=-2iv_{11}.
\end{eqnarray}
The compatibility condition of the equations (\ref{24}) reads  as
\begin{eqnarray}
C_t - G_x + [C, G] =U_{t}-V_{x}+[U,V]= 0\label{38} 
\end{eqnarray}
or in elements   
 \begin{eqnarray}
\kappa_{1t}- \omega_{3x} -\kappa_{2}\omega_{1}+ \tau \omega_2&=&0, \label{39} \\ 
\kappa_{2t}- \omega_{2x} +\kappa_{1}\omega_{1}- \tau \omega_3&=&0, \label{40} \\
\tau_{t}  -    \omega_{1x} - \kappa_{1}\omega_2+\kappa_{2}\omega_{3}&=&0.  \label{41} \end{eqnarray}
We now assume that    
\begin{eqnarray}
\kappa_{1}=i\alpha, \quad \kappa_{2}=-\lambda m, \quad \tau=0 \label{42} 
\end{eqnarray}
and  
\begin{eqnarray}
\omega_{1} & = &-\frac{i\alpha u_{x}}{\lambda},\label{58}\\ 
\omega_{2}&=& \frac{u}{\lambda}+\lambda m(u^{2}-u_{x}^{2}), \label{59} \\
\omega_{3} & = &-2i\alpha[\lambda^{-2}+0.5(u^{2}-u^{2}_{x})].      \label{60}
\end{eqnarray}
Then it is not difficult to verify that  Eqs.(\ref{39})-(\ref{41}) give us the following equations for $m, u$:
\begin{eqnarray}
m_{t}+(m(u^2-u^2_{x}))_{x}+\kappa u_{x}&=&0,\\
 m-u+u_{xx}&=&0.
\end{eqnarray}
It is nothing but the mCHE.  
So, we have  proved  that    the  Lakshmanan (geometrical)  equivalent of the M-CXIIE  is  the mCHE. Finally we note that as follows from (\ref{42}), the  corresponding space curve is with the zero torsion but with the  constant geodesic curvature.

\subsection{Integrable motions of plane  curves}
For the mCHE and that same  for its equivalent counterpart - the M-CXIIE, more naturally  corresponds  the  plane curves than the space curves. For that reason in this subsection, let us consider an  integrable motions of plane  curves induced by the  M-CXIIE.   In this case, Eqs.(\ref{24}) take the following form 
\begin{eqnarray}
\left ( \begin{array}{cc}
{\bf  e}_{1} \\
{\bf  e}_{2} 
\end{array} \right)_{x} = C
\left ( \begin{array}{cc}
{\bf  e}_{1} \\
{\bf  e}_{2} 
\end{array} \right),\quad
\left ( \begin{array}{cc}
{\bf  e}_{1} \\
{\bf  e}_{2} 
\end{array} \right)_{t} = G
\left ( \begin{array}{cc}
{\bf  e}_{1} \\
{\bf  e}_{2} 
\end{array} \right), \label{3.25} 
\end{eqnarray}
where 
\begin{eqnarray}
C=
\left ( \begin{array}{cc}
0   & \kappa_{1}       \\
-\kappa_{1}  & 0    \end{array} \right),\quad G=
\left ( \begin{array}{cc}
0       & \omega_{3}   \\
-\omega_{3} & 0     \end{array} \right).\label{3.26} 
\end{eqnarray}
At the same time, Eqs.(\ref{39})-(\ref{41}) become
   \begin{eqnarray}
\kappa_{1t}- \omega_{3x} =0.  \label{55} \end{eqnarray}
We now assume that    
\begin{eqnarray}
\kappa_{1}=r, \quad \omega_{3} = -r(u^{2}-u^{2}_{x}),      \label{60}
\end{eqnarray}
where
\begin{eqnarray}
r=\sqrt{(u-u_{xx})^{2}+0.5k}.
\end{eqnarray}
Finally Eq.(50) gives
 \begin{eqnarray}
r_{t}+[r(u^{2}-u^{2}_{x})]_{x}=0.      \label{53}
\end{eqnarray}
It is nothing but the mCHE in the conservation law form 
\cite{1310.4011}-\cite{1506.08639}.  
So, again we have  proved  that    the    equivalent counterpart of the M-CXIIE (\ref{6}) is  the mCHE (\ref{53}). Note that in this case, the curve has the zero torsion and normal curvature.

\section{Modified Camassa-Holm equation}

In the previous section, we have proved that the geometrical equivalent of the M-CXIIE is the well-known mCHE. In this section, we give some main informations on the mCHE. The mCHE has the form
\begin{eqnarray}
m_{t}+(m(u^2-u^2_{x}))_{x}+\kappa u_{x}&=&0,\\
 m-u+u_{xx}&=&0.
\end{eqnarray}
The mCHE can be rewritten in the conservation law form as \cite{1310.4011}-\cite{1506.08639}
\begin{eqnarray}
r_{t}+[r(u^2-u^2_{x})]_{x}&=&0,\\
 r-\sqrt{(u-u_{xx})^{2}+\sigma^{2}}&=&0,
\end{eqnarray}
where $2\sigma^{2}=\kappa$. As well-known, the mCHE  is  an integrable nonlinear partial differential equation. Its  LR  read as  \cite{1911.07263}-\cite{1911.12554}
\begin{eqnarray}
\Psi_x&=&U_{3}\Psi,\\
\Psi_t&=&V_{3}\Psi,
\end{eqnarray}
where
\begin{eqnarray}
U_{3}&=&\frac{1}{2}\left(\ba{cc}-\alpha&\lambda m(x,t)\\-\lambda m(x,t)&\alpha\ea\right),\\
V_{3}&=&\left(\ba{cc}\frac{\alpha}{\lambda^2}+\frac{\alpha}{2}(u^2-u^2_{x})&-\frac{u-\alpha u_{x}}{\lambda}-\frac{1}{2}\lambda(u^2-u^2_{x})m\\ \frac{u+\alpha u_x}{\lambda}+\frac{1}{2}\lambda(u^2-u^2_{x})m&-\frac{\alpha}{\lambda^2}-\frac{\alpha}{2}(u^2-u^2_{x})\ea\right)
\end{eqnarray}
with
\begin{eqnarray}
\alpha=\sqrt{1-\frac{1}{2}\kappa \lambda^2}, \quad \phi=u^{2}-u_{x}^{2}.
\end{eqnarray}
Note that
\begin{eqnarray}
V_{3}=-\phi U_{3}+V^{'}_{2}=-\phi U+\left(\ba{cc}\frac{\alpha}{\lambda^2}&-\frac{u-\alpha u_{x}}{\lambda}\\ \frac{u+\alpha u_x}{\lambda}&-\frac{\alpha}{\lambda^2}\ea\right).
\end{eqnarray}
The compatibility condition
\begin{eqnarray}
U_{3t}-V_{3x}+[U_{3},V_{3}]=0
\end{eqnarray}
gives the mCHE (54)-(55). In fact, we have
\begin{eqnarray}
\lambda&:& m_{t}+(m\phi)_{x}+\kappa u_{x}=0,\\
\frac{\alpha}{\lambda}&:& m=u-u_{xx}.
\end{eqnarray}
Note that the mCHE admits at least one reduction. Let   $\kappa=0$,  then the mCHE takes the form
\begin{eqnarray}
m_{t}+(m(u^2-u^2_{x}))_{x}&=&0,\\
 m-u+u_{xx}&=&0.
\end{eqnarray}
Its LR  is given by \cite{1911.07263}-\cite{1911.12554}
\begin{eqnarray}
Z_x&=&U_{4}Z,\\
Z_t&=&V_{4}Z,
\end{eqnarray}
where
\begin{eqnarray}
U_{4}&=&\frac{1}{2}\left(\ba{cc}-1&\lambda m(x,t)\\-\lambda m(x,t)&1\ea\right),\\
V_{4}&=&\left(\ba{cc}\frac{1}{\lambda^2}+\frac{1}{2}(u^2-u^2_{x})&-\frac{u-u_{x}}{\lambda}-\frac{1}{2}\lambda(u^2-u^2_{x})m\\ \frac{u+u_x}{\lambda}+\frac{1}{2}\lambda(u^2-u^2_{x})m&-\frac{1}{\lambda^2}-\frac{1}{2}(u^2-u^2_{x})\ea\right).
\end{eqnarray}

\section{Gauge equivalence between the mCHE and the M-CXIIE}
The M-CXIIE (6) is gauge equivalent to the mCHE (54)-(55). In fact, let us consider 
the gauge transformation
\begin{eqnarray}
\Phi=g^{-1}\Psi,
\end{eqnarray}
where $g=\Psi|_{\lambda=\beta}$. Then the relation between the Lax pairs $U_{1}-V_{1}$ and $U_{3}-V_{3}$ is given by
\begin{eqnarray}
U_{1}=g^{-1}U_{3}g-g^{-1}g_{x}, \quad V_{1}=g^{-1}V_{3}g-g^{-1}g_{t}.
\end{eqnarray}
In the case $\kappa=0$,  the gauge equivalence between the M-CXIE and the mCHE was established in \cite{16}. The gauge equivalence between the mCHE and the M-CXIIE induces some important relations between solutions of these equations. Here we present some of them.
For example, it can be shown that  the solutions $A$ and $m$ is related by the formula
\begin{eqnarray}
tr(A_{x})^{2}=2{\bf A}_{x}^{2}=2(A_{1x}^{2}+A_{2x}^{2}+A_{3x}^{2})=2\beta^{2}m^{2}\end{eqnarray}
or
\begin{eqnarray}
{\bf A}_{x}^{2}=A_{1x}^{2}+A_{2x}^{2}+A_{3x}^{2}=\beta^{2}m^{2}.\end{eqnarray}
Consider the angle parametrization of the spin vector 
\begin{eqnarray}
 A^{+}=\sin\theta e^{i\varphi}, \quad A_{3}=\cos\theta.
\end{eqnarray}
Then from (25) we obtain 
\begin{eqnarray}
{\bf A}^{2}_{x}=\theta_{x}^{2}+\varphi_{x}^{2}\sin^{2}\theta=\beta^{2}m^{2}.\end{eqnarray}
We can  consider the following two particular  cases: $\theta=const$ and $\varphi=const$. In this paper we consider the case $\varphi=const$ and assume that $\beta\in R$. Then Eq.(76) takes the form
\begin{eqnarray}
{\bf A}^{2}_{x}=\theta_{x}^{2}=\beta^{2}m^{2}
\end{eqnarray}
so that
\begin{eqnarray}
\theta_{x}=\pm \beta m.
\end{eqnarray}
We now return to the mCHE (54)-(55).  In terms of $\theta$ it takes the form
\begin{eqnarray}
\theta_{xt}+((u^2-u^2_{x})\theta_{x})_{x}\pm \beta\kappa u_{x}&=&0,\\
 m-u+u_{xx}&=&0
\end{eqnarray}
or
\begin{eqnarray}
\theta_{t}+(u^2-u^2_{x})\theta_{x}\pm \beta\kappa u&=&c,\\
 \theta_{x}\mp \beta(u-u_{xx})&=&0.
\end{eqnarray}

\section{Soliton  solutions of the M-CXIIE}
As the integrable equation, the M-CXIIE has all ingredients of integrable systems like LR, conservation laws, bi-Hamiltonian structure, soliton solutions and so on. In particular, it admits the peakon solutions. Here let us present a one peakon solution of the M-CXIIE. To construct this 1-peakon solution, we use the corresponding 1-peakon solution of the mCHE \cite{0811.2552}. 
\begin{eqnarray}
A=g^{-1}\sigma_{3}g=
\left ( \begin{array}{cc}
A_{3}       & A^{-}  \\
A^{+} & -A_{3}
\end{array} \right),\end{eqnarray}
where
\begin{eqnarray}
g=
\left ( \begin{array}{cc}
g_{1}      & -\bar{g}_{2} \\
g_{2} & \bar{g}_{1}
\end{array} \right), \quad g^{-1}=\frac{1}{\Delta}
\left ( \begin{array}{cc}
\bar{g}_{1}       & \bar{g}_{2}  \\
-g_{2} & g_{1}
\end{array} \right), \quad \Delta=|g_{1}|^{2}+|g_{2}|^{2}.\end{eqnarray}
As result we obtain the following formulas for the components of the spin matrix:
\begin{eqnarray}
A^{+}&=&-\frac{2g_{1}g_{2}}{|g_{1}|^{2}+|g_{2}|^{2}}, \quad A_{3}=\frac{|g_{1}|^{2}-|g_{2}|^{2}}{|g_{1}|^{2}+|g_{2}|^{2}}.  
\end{eqnarray}
To construct the 1-soliton  solution of the M-CXIIE, we consider the following seed solution of the mCHE
\begin{eqnarray}
u=0.  
\end{eqnarray}
Then the equations (58)-(59) have the solutions
\begin{eqnarray}
g_{1}=a_{1}e^{-\theta}, \quad g_{2}=a_{2}e^{\theta},  
\end{eqnarray}
where $a_{j}$ are complex constants and 
\begin{eqnarray}
\theta=\frac{\alpha_{0}}{2}x-\frac{\alpha_{0}}{\beta^{2}}, \quad a_{j}=|a_{j}|e^{i\delta_{j}}, \quad \delta_{j}=consts.  
\end{eqnarray}
 Thus the 1-soliton solution of the M-CXIIE has the form
\begin{eqnarray}
A^{+}&=&-\frac{e^{i(\delta_{1}+\delta_{2})}}{\cosh(\delta-\theta)}, \quad A_{3}=\tanh(\delta-\theta),  
\end{eqnarray}
where $\delta=\ln|\frac{a_{1}}{a_{2}}|$.
\section{Conclusion}
In the paper, one of the GHFE,  namely,  the M-CXIIE  is investigated. The different forms of this equation and its reduction are given. Its  LR is presented. The geometric formulation of this equation is presented. It is also shown that this equation is geometrical and gauge equivalent to the mCHE. 
Finally we note that it is interesting to investigate the surface geometry  of the M-CXXIIE and the mCHE \cite{R13}-\cite{1301.0180}. The M-CXIIE seemingly admits not only soliton but also peakon type solutions. It will be interesting to study the properties of the M-CXIIE and  other integrable GHFE related with Camassa-Holm type equations  in more detail
elsewhere.

\section*{Acknowledgements}
This work was supported  by  the Ministry of Edication  and Science of Kazakhstan under
grants 0118РК00935 and 0118РК00693.

\end{document}